\documentclass[10pt,a4paper]{article}
\usepackage{amssymb}
\usepackage{amsmath}
\usepackage{amsbsy}
\usepackage{amsfonts}
\usepackage{setspace}

\let\ssection=\section
\renewcommand{\section}{\setcounter{equation} {0}\ssection}

\newcommand{\bigspace}{\mspace{10mu}}
\newcommand{\smallspace}{\mspace{5mu}}

\newcommand{\abs}[1]{\mid #1 \mid}

\newcommand{\ket}[1]{\left\vert#1\right>}
\newcommand{\ip}[2]{\left.\left<#1\right\vert#2\right>}

\newcommand{\norm}[1]{\left\Vert#1\right\Vert}

\renewcommand{\vec}[1]{\boldsymbol{#1}}

\newcommand{\limplies}{\Rightarrow}
\newcommand{\liff}{\Leftrightarrow}
\newcommand{\lif}{\Leftarrow}
\newcommand{\ltrue}{\top}
\newcommand{\lfalse}{\bot}
\newcommand{\lsuchthat}{\mid}

\newcommand{\CLtheorem}{{\vdash_{\phantom{}_{CL}} \bigspace }}
\newcommand{\ILtheorem}{{\vdash_{\phantom{}_{IL}} \bigspace }}

\newcommand{\maps}{\longrightarrow}

\newcommand{\monic}{\rightarrowtail}
\newcommand{\epic}{\twoheadrightarrow}
\newcommand{\includes}{\hookrightarrow}
\newcommand{\of}{\circ}
\newcommand{\Sets}{\boldsymbol{Sets}}
\newcommand{\R}{\boldsymbol{R}}
\newcommand{\D}{\boldsymbol{D}}

\newcommand{\C}{\boldsymbol{C}}

\newcommand{\F}{\mathcal{F}}
\newcommand{\G}{\mathcal{G}}
\newcommand{\Z}{\mathcal{Z}}
\newcommand{\B}{\mathcal{B}}

\newcommand{\lA}{\ell A}
\newcommand{\lB}{\ell B}
\newcommand{\lC}{\ell C}
\newcommand{\lX}{\ell X}
\newcommand{\lY}{\ell Y}

\newcommand{\SetsL}{\mathbb{L}}
\newcommand{\SetsR}{\mathbb{R}}
\newcommand{\SetsN}{\mathbb{N}}

\newcommand{\SetsC}{\mathbb{C}}

\newcommand{\SetsQ}{\mathbb{Q}}
\newcommand{\SetsZ}{\mathbb{Z}}
\newcommand{\locisheaves}{\Sets^{\SetsL^{op}}}

\newcommand{\CRings}{\text{C}^{\infty}\text{-Rings}}

\newcommand{\CInfty}{\text{C}^{\infty}}

\newcommand{\topos}{\mathcal{E}}
\newcommand{\truthvalues}{\boldsymbol{\Omega}}
\newcommand{\terminal}{\boldsymbol{1}}
\newcommand{\initial}{\boldsymbol{0}}

\newcommand{\infinitesimals}{\Delta \mspace{-11mu} \Delta}

\newenvironment{proof}
{

    \textbf{Proof}
} {
    \begin{flushright}
    $\blacksquare$
    \end{flushright}
}

\newtheorem{axiom}{Axiom}
\newtheorem{definition}{Definition}

\newtheorem{proposition}{Proposition}
\newtheorem{theorem}[proposition]{Theorem}

\begin{document}

\begin{titlepage}

\title{ \textbf{A Physical Quantum Model in a Smooth Topos} \\
}
\author{John D. Fearns\footnote{email: j.fearns@ic.ac.uk} \\
        \\
        \emph{Blackett Laboratory}\\
        \emph{Imperial College}\\
        \emph{Prince Consort Road}\\
        \emph{London SW7 2BZ}}

\maketitle

\thispagestyle{empty}
\begin{abstract}
We strengthen the case that the new logical perspective afforded by topos theory is suitable to the task of describing the physical world around us. In
exploring some of the aspects of construction of a simple quantum-mechanical system in a mathematical universe different from that represented by set
theory, we show that more thought and a better appreciation of the assumptions going into any mathematical model of the physical world are needed. We
reflect on some of the mathematical consequences of this wider perspective of physics, explaining one interpretation of probabilistic values and numerical
calculations in two mathematical universes governed by the less restrictive intuitionistic logic, and known to support the theory of synthetic
differential geometry.
\end{abstract}

\end{titlepage}

\begin{onehalfspacing}

\section{Introduction}

Frequently in physics, it is the sudden changes in viewpoint that go on to inspire progress. These shifts are usually generalisations not involving total
abandonment of the original ideas. The mathematically motivated shift from the use of the magnetic field to the more commonly discussed vector potential
was one such change. Another was the decision of the earlier natural philosophers to accept the use of the number zero in physics. Such changes in
viewpoint are often resisted by aficionados of the old ways. First, working with a new viewpoint may slow the prediction process of observed phenomena
(ask an engineer to work with the full relativistic vector potential when using magnets). Second, the new viewpoint may predict exactly the same set of
results for any experiment. Thirdly, the new viewpoint requires a greater repertoire of concepts, not all of which immediately sound physically plausible
(how does one observe no bananas? Or empty space?). Those trained to think in the old reliable ways see little point in learning a new way to think, a way
that simply slows one's ability to predict merely the same results. The development of gauge field theory threw new physical fields into the picture.
Fields that were not directly observable. Physicists now treat these fields as more real than their more accessible predecessors. The real gain in having
more than one viewpoint is that a more general feel for physics is obtained. It is this which inspires new ideas. In this paper, we parallel the
representation of the simplest of quantum systems from a wider perspective, that of topos theory as opposed to set theory, and focus on the different
problems encountered along the way, and the extra points of view afforded.

\subsection{Newton and Leibniz}
Some ideas in physics (and maths) collapse during the process of rigorisation, however. One such was Leibniz's viewpoint on calculus. Leibniz
wholeheartedly believed one could manipulate numbers as if there existed infinitesimals. Of course, Newton and Leibniz predated the eventual systematic
use of limits in calculus. Leibniz's differential and integral calculus evolved from a flash of inspiration which recognised the curve as an infinite
succession of infinitely small straight line elements. In his development of the calculus, Leibniz interestingly shifted in his meta-mathematical leanings
away from limits and towards the formal use of arguments involving infinitesimals. Leibniz saw as absolutely fundamental the infinitesimal differential
$dx$ of a quantity $x$. Such a differential would induce the differential $dy(x) = y(x+dx) - y(x)$ on a related quantity $y$. The magnitudes of the
infinitesimals was such that the ratio of these differentials $\frac{dy}{dx}$ gave precisely the tangent to a graph of $y$ against $x$. Leibniz never saw
this ratio as fundamental. He saw it as, begging the pun, derived. With his outlook, the fundamental theorem of calculus was evident when one defined in
analogy to finite sums and differences the infinite sum $\int y dx$: if one takes a finite sequence of numbers such as $1, 3, 8, 5, 6, 8, 4$, and computes
the sum of the differences $(3-1) + (8-3) + (5-8) + (6-5) + (8-6) + (4-8)$ then the result will equal the difference of the first and the last values
$4-1$. In defence of his implicit use of infinitesimal numbers, Leibniz made, in hindsight, some very odd-sounding statements. To his credit though, he
did not try to mask his evident confusion with the pompous language and circular definitions Newton appealed to. Leibniz said~\cite{Newton&Leibniz} that
an infinitesimal was \emph{``less than any given quantity''}, and he compared \emph{``...the neglect of differentials to the work of Archimedes who
``assumed, together with all those following him, that quantities which did not differ by a given quantity were in fact equal.'' ''}. Two months prior to
his death, Leibniz conceded that he \emph{``...did not believe at all that there are magnitudes truly infinite or truly infinitesimal.''}. Nevertheless,
he was firmly of the conviction that infinitesimal reasoning did not collapse into triviality. At first, Leibniz held that infinitesimals were
\emph{unassignable} quantities, suggesting that he believed that infinitesimals had valid existence but that one could never talk about `a given
infinitesimal'. Under pressure from logicians, he retreated to statements that infinitesimals were \emph{qualitative zeroes}. Further backtracking, he
later proposed their existence merely as \emph{auxiliary variables}.

Newton had a rather different perception of the fundamentals of calculus. He left most of the logical and philosophical entanglements associated with
infinitesimal quantities to the intuition of the reader by appealing only to instantaneous velocities. He felt that instantaneous velocity was the
fundamental building block of calculus, and of so obvious a nature that it did not need explaining. Newton became careful later in his life only ever to
proceed in calculations by reference to \emph{ultimate ratios} of these velocities, even though the calculation of his ultimate ratios implicitly used
infinitesimal arguments. Newton's arguments were closer to the eventually rigorous limit-taking foundations developed by Cauchy and others, a century
later.

With the introduction of the limit concept in maths, Newton's and Leibniz's theory found its logical footing rejecting Leibniz's infinitesimals. Ever
since, physicists and mathematicians alike have recognised as logically unsound the use of Leibniz's infinitesimals. Importantly, it might be more to the
point to say that mathematicians and physicists consequently now look down on any possible viewpoint of the world involved with infinitesimals. Today's
mathematical training paints such an unvarying picture of what mathematics is that the overwhelming majority of mathematicians are completely unaware that
exciting and very real alternatives to their methods of thinking exist. The shift in viewpoint parallels the bold step made by Einstein in formulating his
special theory of relativity. Experimentalists reported that the speed of light remained constant from observer to observer, against all the intuitions of
the scientific establishment. It was clear that there is a fundamental relationship between speed on the one hand, and the concepts of time and space on
the other. It also became clear that either the speed of light could vary from observer to observer, preserving the absolute nature of space and time, or
that the speed of light could remain fixed, requiring an unthinkable change towards a more general appreciation of spacetime. The parallel is this: It is
clear that there is a fundamental relationship between maths and physics on the one hand, and the concept of logic on the other. If we wish to insist on
preserving our appreciation of logic, then we equally apply a straightjacket on the imagination of mathematicians and physicists. By subtly generalising
the logic one works with (and the subtle shift is as subtle as the difference between limits of ratios and ratios of limits, but just as non-trivial) it
is entirely possible to consistently model quite beautifully the ideas of Leibniz's calculus. The study of this change will lead to a greater appreciation
of the links between logic and physics, a link which is perhaps more important than either of the subjects alone.

It was not clear to physicists exactly what Einstein was proposing as an alternative when he rejected the idea of absolute space and time, and it took
time to convince his audiences of the self-consistency, and equally importantly, the realism of his picture. Lawvere pioneered as recently as 1970 the
development of topos theory~\cite{Quantifiers&Sheaves}, which allows the perfect modelling of a mathematical universe in which Leibniz's instincts
represent the literal truth, as well as the modelling of countless other universes\footnote{The word `universe', from now on, will be used to refer to a
mathematical universe and not the physical universe.} governed by intuitionistic logic. Intuitionistic logic (IL) is the direct generalisation of
classical logic (CL) where the law of excluded middle is removed as an axiom.

\begin{displaymath} \label{co:EXCLUDEDMIDDLE}
\begin{split}
\CLtheorem \alpha \lor \lnot \alpha \\
\not \ILtheorem \alpha \lor \lnot \alpha
\end{split}
\end{displaymath}

That is to say, it is not generally valid in IL to assume that for any statement $\alpha$, that either $\alpha$ or \emph{not} $\alpha$ is
true\footnote{$\CLtheorem \alpha$ denotes that $\alpha$ is a tautology of propositional classical logic.}. Topos theory models these `mathematical
universes' by using certain categories (toposes) much in the same way that set theory models the `default' mathematical universe using the category of
sets. It turns out that it is precisely the universal validity of the law of excluded middle in CL that blocks any attempts to formulate calculus without
limits. Further, and most interestingly, it is entirely possible to rediscover calculus in universes governed by IL (and modelled by toposes) using
nilpotent infinitesimals, and to go on to recover the theory of differential geometry without limits, in a fashion that would have delighted Leibniz. The
axiomatisation of this theory is called \emph{synthetic differential geometry} (SDG) \cite{Kock,Lavendhomme}. We aim to contrast the classical approach to
physics with a less restrictive one, by taking concrete examples of mathematical universe and physical problem, and approaching the problem inside the
universe. The universe will contain SDG, and the physical problem will be the construction of a simple quantum-mechanical system.

\subsection{Number Systems}
No matter what experiments we perform as physicists, the results of those experiments can always be encoded by a finite series of yes/no data. In any
measurement of distance, we compare the object being measured with \emph{given} distances to obtain a finite amount of greater than/less than comparison
data. One might argue that all possible values of length ratios therefore belong to the rationals $\SetsQ$. Quite clearly this would be obstructive to
progress by mathematical reasoning, and so we take the more mathematically beautiful completion of $\SetsQ$, the reals $\SetsR$. But when we represent
$\SetsR$ by set-theoretic methods we are \emph{already} making assumptions, at least for the following reason. There is more than one way to construct
$\SetsR$, that in CL just happen to yield identical objects. One method defines real numbers by equivalence classes of Cauchy sequences of
rationals\footnote{See any book on analysis for such a definition of the real numbers.}. Physically then, in the context of length, this method might be
interpreted as \emph{defining} length by that invariant which is uniquely determined by an infinite succession of more and more accurate comparison data.
This is an idealist's viewpoint. The second method of defining real numbers is to hold them as equivalent to Dedekind cuts, which split the rationals into
two adjacent collections\footnote{See any book on analysis for such an alternative definition of the real numbers.} (or possibly three if the cut actually
falls on a rational). Physically, the modelling of length by this method might be interpreted as regarding length as something that is already present
before measurements are made. This method would then accept that any comparison measurements would give well-defined greater than/less than results.
Length is \emph{inherent}, and holds just enough information to predict the outcomes of comparisons with yardsticks. This is a realist's viewpoint. The
first method yields, in any universe rich enough for arithmetic\footnote{i.e. A natural numbers object must exist.} to exist, the Cauchy reals $\SetsR_C$.
The second method constructs the Dedekind reals\footnote{When using set theory we drop the subscripts from both sets, because when constructed inside set
theory, $\SetsR_C$ and $\SetsR_D$ are identical.} $\SetsR_D$. Through general logical arguments\footnote{These logical arguments are valid in universes
governed at least by IL, and this of course includes those universes obeying CL. See \cite{Johnstone}.} it is possible to show that the Cauchy reals are a
subobject of the Dedekind reals.

\begin{displaymath}
\SetsR_C \subseteq \SetsR_D
\end{displaymath}

In some universes however, the Cauchies are a proper subobject of the Dedekinds (there are more Dedekind reals than there are Cauchy reals). There are
other methods by which to define a real numbers object, yielding potentially different number systems. From universe to universe, the properties and even
existence of each type of real number object can vary. In any universe containing SDG, there exists an object of \emph{smooth} reals $\R$. This object's
use would appease some of the deepest convictions of those who feel that some aspects of the physical universe should be fundamentally smooth, since it
turns out that all functions from $\R$ to itself \emph{must} be smooth. Again, this is a realist stance, with perhaps more weight placed on assuming $\R$
for its inherent continuum\footnote{In the sense of cardinal arithmetic.} nature than for its relationship to measurements. See the appendix for a brief
mathematical presentation of the some of the number systems mentioned here.

\subsection{Topos Theory}

So as one journeys through different universes, one finds a multitude of `real number' systems, often sharing their existences side by side in the same
universe. As mentioned earlier, topos theory~\cite{Johnstone} is capable of modelling such universes (one universe for each topos), and has presented
logicians and mathematicians with the chance to study them. It should be stressed that a topos is not the same thing as a universe, but a simulation of
it, much as the set $\{ \emptyset, \{\emptyset\}, \{\emptyset, \{\emptyset\}\}, \ldots\}$ simulates the natural numbers. A topos is a certain kind of
category\footnote{See `Toposes' in the appendix.}. The objects (or generalised sets) of the universe it represents are given by the objects of the
category. The maps between objects are given by the arrows of the category. The theorems of the universe are represented by categorical relationships
between objects and arrows, and involve a special \emph{object of truth values}\footnote{This object of truth values is unique up to isomorphism in each
topos, as is the case for the topos of sets and set-functions, for which any two-element set can act as a set of truth values.}.

One example of a topos is the category of sets and set-functions\footnote{Although we refer to \emph{the} category of sets and set-functions, there is
more than one. This fact does not alter our presentation of topos theory.}. Its objects are represented by sets, its maps are represented by
set-functions, and the theorems of the mathematical universe it represents can be formulated as relationships between various sets, functions and the
special set $\{ true, false \}$. The category of sheaves over a topological space\footnote{See `Sheaf Categories as Toposes' in the appendix.} is always a
topos~\cite{Sheaves}. Objects in the universe represented by the topos are given by the sheaves. Maps in the universe are represented by the sheaf
morphisms.

The study of toposes gives rise to slightly more general interpretations of the basic logical manipulations of sets and their elements, uniquely lifting
all the familiar set-theoretic notation ($\{, \}, \forall, \exists, \in, f:A \maps B$, etc) to the universes modelled by them. Far from being `unreal',
these universes are highly self-consistent and structurally rich environments in which to \emph{think}\footnote{In the sense of a deductive logic.}. The
theorems of one topos need not hold in another (nor the objects and maps to which they refer). When working with others to establish shared mathematical
knowledge one can either fix a particular topos in which to exclusively work forever, or one can work in many, keeping in mind what topos theory says
about how the universes are related. The latter is evidently more general. The former is the path humans have taken until now. It is entirely reasonable
that alien civilisations have in parallel unwittingly chosen other default toposes in their early mathematical development. Such a different choice of
path would lead to different mathematics, and a different appreciation of physics.

For an excellent introduction to topos theory see \cite{Goldblatt}. For those familiar with category theory and $\CRings$ see the appendix for a very
brief introduction to the (complicated) use of toposes in modelling universes containing synthetic differential geometry.

\subsection{Synthetic Differential Geometry}
As probably guessed by now, it is precisely the universes containing SDG that validate Leibniz's thinking. SDG is a theory that rests on the existence and
a few axiomatic properties of the smooth reals $\R$, and first-order intuitionistic logic. The axiomatic development of SDG, as well as initial searches
for models of SDG (IL universes in which the axioms are validated) were pioneered by A.~Kock. The most complete survey of toposes modelling SDG is `Models
of Smooth Infinitesimal Analysis' by I.~Moerdijk and G.E.~Reyes~\cite{Models}. Described within are some of the so called well-adapted models of SDG,
universes in which there is a way to faithfully rebuild a large category of classical manifolds using $\R$ instead of $\SetsR$. We assume all toposes
mentioned in this paper to be well-adapted models of SDG unless stated otherwise. The appendix contains a reference list of relevant notation from
axiomatic SDG.

The most important subobject of $\R$ to differential geometry is the object of first-order infinitesimals $\D$.
\begin{displaymath}
\D := \{d\in \R|d^2=0\} \mspace{20mu}
\end{displaymath}
This object's properties usefully illustrate the main differences between CL and IL. In IL, the logical existence of an element $x$ of an object $X$
($\exists x \in X$) is a notion that expresses the fact that the element can in principle be explicitly pinpointed and assigned a name (e.g. the zero of
$\R$ exists in $\R$). Most of the difficulty in transferring one's thought patterns up from CL to IL is the conquest of the ingrained assumption that one
can in principle imagine in crystal clarity each and every one of the elements of an object $X$ \emph{separately}. This is not a guaranteed property of
every object in an IL universe. This does not preclude a large class (called the \emph{decidable} objects) of the objects in an IL universe from having
this property. All sets in set theory are decidable.
\begin{definition}
An object $X$ is decidable when
\begin{displaymath}
(\forall x,y \in X) ((x=y) \lor (x \neq y))
\end{displaymath}
\end{definition}
\textit{e.g. $\SetsR_C$ is decidable, and $\R$ is not decidable}.
\newline

The following logical statements concerning $\D$ should be considered
\begin{displaymath}
\begin{split}
\lnot (\forall d \in \D)(d=0)\\
\lnot (\exists d \in \D)(d \neq 0)
\end{split}
\end{displaymath}
So the object $\D$ is definitely not equal to the object $\{0\}$, but it is not possible to pinpoint any elements of $\D$ other than zero. The elements of
$\D$ are all indistinguishable from each other, but the important feature of IL is that indistinguishability of elements does not imply equality of
elements. As a direct consequence of
\begin{displaymath}
\not \ILtheorem \alpha \lor \lnot \alpha
\end{displaymath}
We have that
\begin{displaymath}
\not \ILtheorem \lnot \lnot \alpha \implies \alpha
\end{displaymath}
i.e. $\alpha$ being indistinguishable from true (not untrue) does not necessarily imply $\alpha$ is true. This behaviour of logical statements parallels
the behaviour of elements of objects in universes governed by IL. The word `indistinguishable' should be understood in IL to refer to the double negation
of an equality, and not to the CL consequence of the actual equality.

In SDG $\R$ obeys the axioms listed in the appendix, and in particular the Kock-Lawvere axiom \setcounter{axiom}{5}
\begin{axiom}[Kock-Lawvere] \label{axiom:KL}
For each $f:\D \maps \R$, there exists a unique $b \in \R$, such that for every $d \in \D$
\begin{displaymath}
f(d)=f(0)+d.b
\end{displaymath}
\end{axiom}

Through direct use of this axiom is defined the derivative of a function at all points of $\R$. A generalisation of this axiom extending this
microlinearity of functions to greater orders and dimensions is also assumed. Further, in SDG there exists an integration axiom guaranteeing for each real
function $g$ on $\R$ the unique (up to the addition of a constant) existence of an antiderivative $G$. It is necessarily true that all functions $\R
\xrightarrow{g} \R$ are infinitely differentiable. More features of SDG follow as and when we need to talk about them.

\subsection{Representing Physics using other Universes}
The representation of physics by the mathematics of one universe or another will naturally vary. Excellent work in the representation of classical physics
in universes containing SDG has been done~\cite{SHM}, as well as some work on the reconstruction of general relativity in the same context~\cite{Guts}. It
is our aim to elucidate the methods used to construct basic quantum mechanics in this context, and to contrast and compare them with the mainstream
treatment in set theory, from the perspective of physics. The idea, from the outset, is that the physics being represented is the same physics whether one
universe or another is used to study it. In the course of development, it may be necessary to reject certain universes as unsuitable, or to modify one's
appreciation of physics itself, hopefully the latter. In this paper we examine parts of the construction of a simple quantum-mechanical system, discuss
any immediate consequences, and suggest areas for further development. The system for consideration is the spin state-space of a pair of
spin-$\frac{1}{2}$ particles, chosen because this is the simplest possible compound quantum state-space there is. It should be stressed that most of the
mathematical machinery used in physics has not been built in other universes. Such concepts as vector spaces, inner-products, Hilbert spaces, Lie groups,
representations and suchlike depend on the number systems one bases them on, and consequently are largely unstudied. We must continue with a degree of
care and an awareness of the potentially important assumptions one makes along the way. That said, it would be impossible to make progress embracing
fastidious drudgery.

\section{The Maths of Quantum Mechanics}
We will use the term \emph{classical} to refer to set-theoretic methods, and not to pre-quantum-physics. Classically, quantum state-spaces have been built
as representations of the Lie algebras of symmetry Lie groups. Some of the most simple representations are of the Lie algebra of the rotation group SO(3).
If, however, space is modelled using $\R^3$ instead of $\SetsR^3$, a corresponding modification to our representations should also be made. The
motivations are mostly mathematical. As an example, the mixing of the use of $\R$ in the definition of a rotation group, and $\SetsR_C$ in a linear
representation of that group are incompatible for all but the trivial representation. We prove this by considering the necessary Lie algebra homomorphisms
involved. Any Lie algebra homomorphism from an $\R$-algebra to an $\SetsR_C$-algebra requires the existence of a ring homomorphism from $\R$ to
$\SetsR_C$. The following proof makes use of the injective ring homomorphism $i: \SetsR_C \maps \R$, which is present in any topos modelling SDG.

\begin{proposition}
No non-trivial ring homomorphism $\theta: \R \maps \SetsR_C$ exists.
\end{proposition}
\begin{proof}
Suppose some such $\theta$ exists. Since $\SetsR_C$ is a subring of $\R$, we require that the restriction to $\SetsR_C$, $\theta \mid_{\SetsR_C}$, is
itself the non-trivial ring homomorphism of $\SetsR_C$, the identity map.
\begin{displaymath}
(\forall x \in \SetsR_C)(\theta(x)=x)
\end{displaymath}
Taking any two indistinguishable elements $x,y$ of $\R$ (we are using implicit naming here, not explicit naming: this is allowed in IL) we have that
$\theta(x)$ and $\theta(y)$ are also indistinguishable (otherwise we have a method of distinguishing $x$ and $y$).
\begin{displaymath}
(\forall x,y \in \R)(\lnot (x \neq y) \limplies \lnot (\theta(x) \neq \theta(y)))
\end{displaymath}
$\SetsR_C$ is a decidable object in any topos.
\begin{displaymath}
(\forall x,y \in \SetsR_C)((x=y) \lor (x \neq y))
\end{displaymath}
Hence for indistinguishable $x$ and $y$, it must be that $\theta(x) = \theta(y)$. Noting that $(\forall d \in \D)(\lnot (d \neq 0))$, we have that
\begin{displaymath}
(\forall x \in \R)(\forall d \in \D)(\theta(x+d) = \theta(x))
\end{displaymath}
Viewing the map $\theta$ as a map from $\R$ to $\R$ (by composing with $i$), we have $\frac{d\theta}{dx} \equiv 0$, and by the fundamental theorem of
calculus and the observation that $\theta(0)=0$, we require that $(\forall x)(\theta(x) = 0)$, contradicting the initial supposition.
\end{proof}

Mathematically, the unsuitability of the Cauchy reals for the field over which representations are made rests on the ability of non-constant maps from
$\R$ to \emph{any} decidable object $X$ (including $\SetsR_C$) to be viewed as maps that split $\R$ into a union of non-intersecting
subobjects\footnote{In set theory, the analogue is that any function $f:\SetsR \maps X$ divides $\SetsR$ into a union of equivalence classes given by the
equivalence relation $x \sim y \iff f(x)=f(y)$.}. Any such split is logically impossible for the smooth real line because of its very strong continuity
properties. It simply cannot be considered as a union of disjoint parts. If we could split the smooth reals into two distinct subobjects, we could use
these subobjects to construct a function not obeying the Kock-Lawvere axiom. So $\R$ is unlike any set of set theory. It should, however, be entirely
possible to build representations of groups with $\R$-Lie-Algebras using any real number system $\SetsR_i$ that admits a non-trivial ring homomorphism $\R
\maps \SetsR_i$.

This all suggests Hilbert spaces defined over the field\footnote{`Field' takes on a modified meaning in SDG} $\C$ should be built and used to represent
physical states in quantum mechanics\footnote{Here $\C$ is the smooth complex field defined as $\R \times \R$.}. So until a motivation not to do so
exists, we will use $\R$ for our group representations, and consider the ramifications for the concept of probability later on.

\subsection{Vector Spaces over $\R$}
In SDG a tangent vector on a microlinear space\footnote{A microlinear space $M$ is defined in SDG as an object having in a certain sense similar
infinitesimal properties to $\R$ itself. This can be thought of as assuming that $M$ behaves as if it is capable of supporting an infinitesimally
extensive coordinate chart at each of its points. See \cite{Lavendhomme}.} $M$ is simply defined as a map
\begin{displaymath}
t : \D \maps M
\end{displaymath}
This definition is along the same lines as the classical definition, but is far simpler, and neatly encapsulates the notion of `vector' as an
infinitesimal line embedded in the space $M$. This definition of tangent vector motivated a definition of vector space as a Euclidean $\R$-module (see
Lavendhomme's introductory book on synthetic differential geometry, \cite{Lavendhomme}).
\begin{definition}
A Euclidean $\R$-module is an object $V$ equipped with an $\R$-module structure satisfying the Kock-Lawvere axiom
\begin{displaymath}
(\forall t \in V^{\D})(\exists !(x,y) \in V \times V)(\forall d \in \D)(t(d) = x + dy)
\end{displaymath}
\end{definition}
\textit{e.g. all spaces of the form $\R^X$, with the obvious $\R$-module structure}.
\newline

The notation $A^B$ in topos theory\footnote{It is a requirement of a topos that $A^B$ exists for all pairs $A,B$.} represents the object of maps from
$B$ to $A$. This definition of vector space looks slightly more restrictive than the classical counterpart. The requirement of satisfying the
Kock-Lawvere axiom imposes the condition that a vector space be isomorphic to its own tangent space at the origin. The classical definition of a real
vector space requires that $V$ simply be an $\SetsR$-module. However, such a definition of vector spaces in IL would include, for example, the object
$\D_{\infty}$ of all nilpotent infinitesimals. $T_0(\D_{\infty})$, the tangent space to $\D_{\infty}$ at $0$ is isomorphic to $\R$, which is `larger'
than $\D_{\infty}$ (this fact is related to the question of the existence of exponential maps from Lie Algebras to Lie groups. In this case, the
exponential map from the Lie algebra $\R$ of the Lie group $\D_{\infty}$ does not exist). e.g. all spaces of the form $\R^{X}$, with the obvious
$\R$-module structure.
\subsection{Finite Summation}
The finite cardinal objects in any topos are the objects $[n] := \{m \in \SetsN \lsuchthat m < n\}$. They are the same thing as the objects
$\{0,1,\ldots,n-1\}$. Topos-theorists then define \emph{finiteness} as follows.
\begin{definition}
An object $F$ is finite when there exists a surjection\footnote{In topos theory, a map $s:A \maps B$ is surjective iff for every pair of maps $f_1, f_2: B
\maps C$ such that $f_1 \of s = f_2 \of s$ it is true that $f_1 = f_2$. Note that this categorises the surjections of set theory also.} $[n] \epic F$ for
some $n \in \SetsN$.
\end{definition}

It is interesting to note that a finite object is not necessarily the same thing as a finite cardinal, unless that object is also decidable. This is true
in all universes modelled by topos theory. So it may not at first be obvious what one means by finite summation. The sum of a finite sequence of values is
not well-defined unless that finite sequence is also a decidable sequence. Noting that $\SetsN$ is decidable, taking any sequence of real numbers $a \in
\R^{\SetsN}$, the sum $\sum_{i \leq n} a_i$ is defined by recursion\footnote{The natural numbers object $\SetsN$ of a topos is defined precisely to allow
the recursive definition of functions.}: $\sum_{i \leq 0} a_i = a_0$, $\sum_{i \leq n+1} a_i$ =  $a_{n+1} + \sum_{i \leq n} a_i$. The variant $\sum_{1
\leq i \leq n}$ or $\sum_{i=1}^{n}$ can also be defined using this recursive definition, where the last two symbols are interchangeable. This motivates
the following natural definition, included mostly to provide a necessary and more specific meaning to a very familiar word.
\begin{definition}
A map $C \xrightarrow{a} A$ is a \emph{collection (of elements of A)} when $C$ is decidable.
\end{definition}
Then any finite collection of elements $\{v_1,v_2,\ldots,v_n\}$ of a vector space may be summed to $\sum_{i=1}^n v_i$.

\subsection{Bases}
Set-theoretically, if the free real vector spaces over the sets $X$ and $Y$ are isomorphic, then $X \cong Y$.
\begin{displaymath}
\SetsR^X \cong \SetsR^Y \limplies X \cong Y
\end{displaymath}
Consequently if $X$ is finite, then so is $Y$. The situation in intuitionistic universes differs though. A vector space of the form $\R^I$ may also be of
the form $\R^J$ for some $J \not \cong I$. As an example, take the vector spaces $\R^{\D}$ and $\R^{[2]}$. From the topos-theoretic definition of
\emph{finite object} it is clear that the object $\D$ of square-zero reals is \emph{not} a finite object (any map from any $[n]$ to $\D$ must map all
elements of $[n]$ to zero, otherwise we have a method to explicitly name an element in $\D$ other than zero, which we know is disallowed). It is also
clear that the object $[2]$ \emph{is} a finite object. Yet the two vector spaces are isomorphic by the Kock-Lawvere axiom.

Classically, the notion of basis is well understood. For finite-dimensional Hilbert spaces, Hamel bases and Hilbert-space bases coincide, but even in this
case the example above demonstrates that the lifting of either concept of `basis' to other universes is tricky. The notion of \emph{basis set} needs to be
made more specific to be unambiguous, when one leaves set theory. This is fundamentally due to the added structure found in the objects and morphisms of a
topos, that is missing from set theory. Classically, when we select a \emph{basis set} we pay no attention to the structure of this set, because
categorically, there is no structure to be elicited (no set has categorical structure beyond its collection of distinguishable elements). One rather more
well-behaved type of exponent defines the following type of vector space.

\begin{definition}
A vector space $\R^X$ has \emph{cardinal support} if it has a finite cardinal exponent $X \cong [n]$ (for $n \in \SetsN$).
\end{definition}
e.g. $\R, \R^n, \R^{[n]}$

\begin{proposition} \label{prop:UNIQUEINDEX}
A vector space with finite cardinal support has a unique finite cardinal exponent $[n]$.
\end{proposition}

\begin{proof}
Let $\R^{[n]} = \R^{[m]}$. Note that $n < m \limplies \R^{[n]} \subset \R^{[m]}$. $\SetsN$ being linearly ordered, we
have $(\forall n,m \in \SetsN) ((n<m) \lor (n=m) \lor (n>m))$. $\SetsN$ is also decidable, hence $(n \neq m) \liff
((n<m) \lor (n>m))$. We may now prove by contradiction.
\end{proof}
We note that the decidability of the exponent object plays a crucial role in its uniqueness, mathematically suggesting that the definition of basis should
involve the use of decidable subobjects of a vector space, or \emph{collections} of vectors.

Turning to the physical motivations in the use of bases, we note that most vector spaces in physics are tied strongly to very physical concepts, such as
space, or state. Whenever we choose a basis for such a vector space, we implicitly provide a certain and \emph{perceptibly different} physical meaning to
each of the basis elements (e.g. spin up and spin down), and are consequently interested in finding the `amount' to which a vector `has' that meaning.
This suggests a physical reason why bases should be defined by decidable subobjects of a vector space. In classical physics, this would amount to assuming
a decidable phase space. Of course, the restriction of the modelling of phase spaces to using decidable objects is an idealist attitude, in the same vein
that an idealist might want to model space using $\SetsR$ or even $\SetsQ$, and does not take maximum advantage of all intuitionistic availabilities.
There would be equally valid arguments for the explicit use of undecidable exponent objects (in the quantum mechanics of wavefunctions). It will be
possible to consider the quantum mechanics of undecidable systems, but that is left for another paper, since in this paper we ultimately wish to consider
only the distinguishable spin states of a spin-$\frac{1}{2}$ particle. We suggest the following natural (if slightly more specific than for set theory)
definitions for vector spaces.

\begin{definition}
For a vector space V, a \emph{finite linear combination} of vectors is a sum $\sum_{i=1}^{n} v_i$ where the $v_i$ are a finite collection of vectors $[n]
\xrightarrow{v} V$. Note that the $v_i$ must be a \emph{collection} (and not merely finitely-indexed) in order to well-define the sum $\sum_{i} v_i$.
\end{definition}

\begin{definition}
For a vector space V, a collection of vectors $L \xrightarrow{e} V$ is \emph{linearly independent} if for all finite linear combinations of vectors in
$L$, $(\sum_{i=1}^{n} v^ie_i = 0) \limplies (\forall i)(v^i = 0)$. (Note that any subobject of $L$ defines a restricted collection, allowing finite linear
combinations to be well-defined).
\end{definition}

\begin{proposition} \label{prop:LI}
A linearly independent collection of vectors $L \xrightarrow{e} V$ is a proper subobject of $V$.
\end{proposition}

\begin{proof}
To show that $e$ is injective (categorically monic), take $e(m) = e(m')$. Suppose that $m \neq m'$. Note that $\lambda e(m) - \mu e(m') = 0$ requires
$\lambda = \mu = 0$, contradicting the supposition (via the real line axiom $1 \neq 0$). Proof by contradiction is then allowed to proceed by the
decidability of $L$.
\end{proof}
This may sound obvious to a set-theorist, but it is a fact that heavily relies upon the decidability of $L$. A more general approach to summation might
not yield a similar result, requiring summation and linear independence to become more abstract notions.

\begin{proposition} \label{prop:NONZERO}
All vectors in a linearly independent collection are distinct from the zero vector (as well as from each other).
\end{proposition}

\begin{proof}
Suppose that some $e_i = 0$. Then $1.e_i = 0.e_i$ contradicting linear independence by the real line axiom $1 \neq 0$. Hence $(\forall i)(e_i \neq 0)$.
Note that this proof is intuitionistically valid, as we have not appealed to the excluded middle, $(e_i = 0 \lor e_i \neq 0)$. We have merely shown that
$(e_i=0) \limplies \lfalse$, which is the logical definition of $e_i \neq 0$. That the elements are distinct from each other is shown by the decidability
of $L$, and proposition (\ref{prop:LI}).
\end{proof}
Again, a result confirming the subobject nature of linearly independent collections of vectors, which relies upon decidability.

\begin{definition}
A collection of vectors $S \xrightarrow{s} V$ \emph{spans} $V$ if every vector in $V$ can be written as a finite linear combination of vectors in the
collection.
\end{definition}

\begin{definition}
A \emph{basis} for a vector space $V$ is a collection of vectors $E \stackrel{e}{\monic} V$ that
\newline (i) is linearly independent \newline (ii) spans $V$
\end{definition}

\begin{proposition}
Given a basis for a vector space $V$, for each vector $v$, let $v^i$ be the coefficients in the linear combination of basis vectors summing to $v$. Then
$(v=w) \liff (\forall i)(v^i = w^i)$.
\end{proposition}

\begin{proof}
We may prove this constructively, in exactly the same manner as for set theory: $\lif$ is obvious. $\limplies$ can be proved by considering the vector
$v-w = 0$, whose coefficients are given by $v^i-w^i$, each of which must equal zero by linear independence of the basis.
\end{proof}

\begin{theorem}
Let $V$ be a vector space with a finite basis $[n] \stackrel{e}{\includes} V$. Then any other basis $[m] \stackrel{e'}{\includes} V$ has the same exponent
object, i.e. $n=m$. We call $n \in \SetsN$ the \emph{dimension} of the finite-dimensional vector space $V$.
\end{theorem}

\begin{proof}
We construct the vector space isomorphisms $\R^{[n]} \cong V \cong \R^{[m]}$ and appeal to proposition (\ref{prop:UNIQUEINDEX}). Concentrating on the case
$V \cong \R^{[n]}$ we define
\begin{displaymath}
\begin{array}{ll}
\theta: V \leftrightarrow \R^{[n]} & \text{where}\\
\theta(v)(i) := v^i & \text{and}\\
\theta^{-1}(\tilde{v}) := \sum_{i=1}^{n}\tilde{v}(i)e_i
\end{array}
\end{displaymath}

This is evidently a vector space isomorphism. The similar result for the second basis establishes $\R^{[n]} = \R^{[m]}$ and hence that $m=n$.
\end{proof}

\subsection{Linear Maps}
The ability to represent vectors of finite-dimensional vector spaces by lists of their coefficients extends, as expected, to the representation of linear
maps by matrices.

Let $\theta: V \maps W$ be a linear map between two finite-dimensional vector spaces with respective bases $\{e_j\}$ and $\{\tilde{e}_i\}$ . Then
\begin{displaymath}
\theta(\sum_{j=1}^{m}v^je_j) = \sum_{j=1}^{m}v^j\theta(e_j) = \sum_{i=1}^{n}\sum_{j=1}^{m}\theta^{ij}v^j\tilde{e}_i \text{ for some unique smooth-real
matrix } \theta^{ij}
\end{displaymath}

Again, this is crucially due to the finiteness and decidability requirements in the definition of a basis, which allows us to use the unique
representation of vectors to determine each row of the matrix in turn. We are effectively permitted the use of classical logic in arguments where lists of
coefficients need to be compared and manipulated. Any properties of vector spaces over $\R$ that are different from the corresponding classical vector
spaces (over $\SetsR$ or its equivalent $\SetsR_C$ in any topos possessing it) are therefore due to the nature of the coefficients themselves, rather than
the way in which the lists of coefficients are handled. This need not hold true of vector spaces of the form $\R^F$ where $F$ is a finite object for which
decidability has not been established.\newline

We may define the tensor product and tensor sum of vector spaces in the usual categorical manner\footnote{The tensor product and tensor sum of a pair of
vector spaces are precisely the categorical limit and co-limit of the vector spaces in the category consisting of all vector spaces and linear maps.}. Due
to the following two properties of finite cardinal objects, we observe the usual dimensional relationships between vector spaces and their tensor products
and sums. In any topos with a natural numbers object, for all $n,m \in \SetsN$

\begin{displaymath}
\textsl{Cartesian product and disjoint union} \left\{ \begin{array}{lcl}
\left[ n \right] \times [m] &= &[n \times m] \\
\left[ n \right] + [m] &= &[n + m]\\
\end{array} \right\} \textsl{Algebraic multiplication and addition}
\end{displaymath}

Since $\R^{[n]} \otimes \R^{[m]}$ is isomorphic as a vector space object to $\R^{[n] \times [m]}$, the dimension of the the tensor product is equal to the
product of dimensions. The similar result for tensor sums also stands. We are then allowed to consider the canonical bases $\big\{
\ket{\xi_i}\otimes\ket{\eta_j}\big\}$ of the tensor product of two vector spaces with respective bases $\big\{\ket{\xi_i}\big\}$ and
$\big\{\ket{\eta_j}\big\}$, following the classical result.

\subsection{Inner-products and Norms}
In this section, the axiomatics of $\R$ play a more prominent role. From now on we also work with complex vector spaces. We define the complex numbers
object as the usual $\C \cong \R^2$. Following the classical definition of an inner-product we define
\begin{definition}
An \emph{inner-product}, $\ip{\smallspace}{\smallspace}$, on a finite-dimensional vector space $V$ is a strictly positive, symmetric, bilinear functional
$p: V \times V \maps \C$.
\end{definition}
We take \emph{strictly positive} to mean that $(x \neq 0) \limplies (\ip{x}{x} > 0)$. Axiom (\ref{axiom:A3}) relates positivity and invertibility, so we
have that
\begin{proposition}
The only way any inner-product can exist on $\R$ is if $\R$ is field in the following sense
\begin{displaymath}
(\forall x \in \R)(x \neq 0 \limplies x \in U(\R))
\end{displaymath}
where $U(\R) = \{x \in \R \lsuchthat \exists x^{-1}\}$.
\end{proposition}
\begin{proof}
Take an inner-product $\ip{\smallspace}{\smallspace}$. $(\forall x,y \in \R)(\ip{x}{y} = \alpha xy)$ where $\alpha = \ip{1}{1}$ is positive. For $x=y \neq
0$ our inner-product requires $\alpha x^2 > 0$. Axiom (\ref{axiom:A3}) then succeeds with
\begin{displaymath}
(\alpha x^2 > 0) \limplies (x^2 >0) \limplies (x > 0) \limplies (x \in U(\R))
\end{displaymath}
\end{proof}
We take the slightly stronger axiom suggested by Moerdijk and Reyes~\cite{Models}. \setcounter{axiom}{3}
\begin{axiom}[Field]
\begin{displaymath}
(\forall x_1,\ldots,x_n \in \R) \Big( \lnot(x_1=0 \land \ldots \land x_n=0) \limplies (x_1 \in U(\R) \lor \ldots \lor x_n \in U(\R))\Big)
\end{displaymath}
\end{axiom}
Without $\R$ obeying at least the weaker field axiom, we would find it more difficult to deal with (and define) inner-product spaces. The canonical
product $\sum_{i=1}^{n}\overline{x^i}y^i$ of two vectors in $\C^n$ would fail to be an inner-product. A field axiom acts as a bridge between the algebraic
and subobject properties of a ring. An inner-product is something that rests on this bridge, and tells us that any non-zero vector can be stretched by a
unique finite amount, to meet the unit sphere. This axiom is relevant to quantum theory, as shall be seen when self-adjoint operators are briefly
considered. Luckily, many models of SDG so far discovered show $\R$ to be such a field.

We may define orthonormal bases in the usual manner.
\begin{proposition}
The Gram-Schmidt process for generating orthonormal bases is valid for finite-dimensional inner-product spaces.
\end{proposition}
Classically, a recursive method is used to prove the theorem for all dimensions $n \in \SetsN$. Here we emulate the first step of the classical proof,
taking note of which properties of $\R$ are essential. The complete classical proof is entirely valid here.
\begin{proof}
Take any basis $\{b_1,\ldots,b_n\}$ for the finite-dimensional inner-product space. By proposition (\ref{prop:NONZERO}), $b_1 \neq 0$. By positivity and
symmetry of the inner-product, $\ip{b_1}{b_1}$ is real and positive. By axiom ({\ref{axiom:A3}}), $\norm{b_1}$ exists and is also positive, and
$\norm{b_1}^{-1}$ exists. Defining $e_1 := \frac{b_1}{\norm{b_1}}$ we have obtained the first vector of our orthonormal basis $\{e_1,\ldots,e_n\}$.
\end{proof}
For the completion of the proof, there is also a reliance on the decidable and finite nature of $[n]$. The validity of Gram-Schmidt allows us to pick an
orthonormal basis for any finite-dimensional inner-product space, which is required in the labelling of spin-states.

\subsection{Hermitian Operators}
\begin{definition}
A linear operator $A$ is \emph{hermitian} with respect to the inner-product $\ip{\smallspace}{\smallspace}$ when for all $x$, $\ip{Ax}{x} \in \R$.
\end{definition}
It is easy to see that any eigenvalues of a hermitian operator must belong to $\R$, which is the most immediately relevant property of hermitian operators
for quantum mechanics.

As for previous sections, most classical results transfer directly to the alternative universes. As an example that follows the classical proof,
\begin{proposition}
Distinguishable eigenvalues of a hermitian operator yield orthogonal eigenspaces.
\end{proposition}
\begin{proof}
Take the operator, eigenvectors and eigenvalues
\begin{displaymath}
    \left. \begin{array}{c} A\ket{\lambda} = \lambda \ket{\lambda} \\
    A\ket{\mu} = \mu \ket{\mu} \\ \end{array} \right\}
    \lambda \neq \mu
\end{displaymath}
Then
\begin{displaymath}
\lambda - \mu \neq 0 \text{ and } (\lambda - \mu)\ip{\lambda}{\mu} = 0
\end{displaymath}
hence, by the field axiom (\ref{axiom:FIELD})
\begin{displaymath}
\ip{\lambda}{\mu} = 0
\end{displaymath}
\end{proof}

The field axiom of $\R$ comes into play here. Without the field axiom holding, one cannot generally show the orthogonality of the eigenstates of two
different eigenvalues, although one can prove (using axiom (\ref{axiom:A3})) that for two different eigenvalues $\lambda$ and $\mu$, their eigenvectors'
inner-product must not be invertible as a complex number, $\lambda \neq \mu \limplies \ip{\lambda}{\mu} \not \in U(\C)$.

We have defined and proven some elementary properties of bases that are required to form the quantum-mechanical representation of the next section.

\section{A Physical Example}

\subsection{Hilbert Space Representations}

Our physical system will be an approximation of the spin-spin interaction of two electrons in a helium atom. As argued earlier, the assumption that space
is modelled by $\R^3$ naturally leads to quantum representations being defined over the field $\C$ rather than over the field $\SetsC$. We assumed that
the two-dimensional complex vector space $\C^2$ is suitable as a Hilbert space object for the spin of one electron. However, we make no attempt to define
the term `Hilbert space object' in any of the toposes modelling SDG. In set theory, a finite-dimensional complex Hilbert space is precisely the same thing
as a finite-dimensional complex inner-product space. $\C^2$ is a finite-dimensional complex inner-product space. We may not, however, reasonably assume
$\C^2$ to be the analogue of a Hilbert space, for it is well known by synthetic differential geometers that $\R$ is not a Cauchy-complete space. It is
interesting that the incompleteness of $\R$ does not imply that a non-convergent Cauchy sequence exists. Further, it is possible to show that one cannot
find any non-convergent Cauchy sequences. This perhaps counterintuitive situation is permitted by the lack of validity of the classical axiom $\lnot
\big(\forall x \phi(x)\big) \limplies \big(\exists x \lnot \phi(x)\big)$, which is related to the (missing) law of excluded middle. We continue in the
assumption that all quantum state-spaces referred to from now on are finite-dimensional.

Marching on towards quantum physics, we assume a well-defined notion of Lie Group (that has a local structure similar to that of $\R$) beyond that of
microlinear group (which has merely an infinitesimal structure similar to that of $\R$). Discussions of Lie groups in SDG can be found in \cite{Goldblatt}
and \cite{LieIntegration}, though it should be noted that their definitions of Lie group differ. We will accept the stronger definition (Kock). All the
unitary groups $U(n)$ are then still Lie groups in any of the usual well-adapted models of SDG.

For each electron, we use an internal two-dimensional representation of the familiar Lie Algebra of $SO(3)$, and an analogue of the usual Hamiltonian in
terms of these generators. Of course, when we say \emph{algebra}, we mean an algebra over the field $\R$, and not over the field $\SetsR$.
\subsection{SO(3)}
We define the group $SO(3)$ as the group of $3 \times 3$ unit determinant matrices preserving the natural inner-product on $\R^3$.
\begin{displaymath}
SO(3) := \left\{ U \in \R^{3 \times 3} \lsuchthat UU^T = 1, det(U) = 1 \right\}
\end{displaymath}
The group, as defined, is a subobject of $\R^9$ and is hence a \emph{microlinear group}\footnote{As such, it has an easily defined Lie Algebra, see
\cite{Lavendhomme}.} in SDG. Its Lie algebra is given by the tangent space at the identity
\begin{displaymath}
\mathcal{L}(SO(3)) := \left\{ t:\D \maps SO(3) \lsuchthat t(0) = 1 \right\}
\end{displaymath}
We know that since $\R^9$ is a vector space, any tangent vector $t$ at the identity element is of the form
\begin{displaymath}
t(d) = 1 + X.d
\end{displaymath}
for some unique matrix $X$. For $t$ to be a tangent vector belonging to $SO(3)$ (and not just to $\R^9$) we must have
\begin{displaymath}
(\forall d \in \D)(t(d)t(d)^T = 1)
\end{displaymath}
requiring
\begin{displaymath}
(\forall d \in \D)(1 + d(X + X^T) = 1)
\end{displaymath}
and hence that $X$, as in the classical case, must be an antisymmetric matrix. The standard Lie algebra basis can then be chosen.
\begin{displaymath}
\begin{array}{ccc}
{L_1 = \left(
\begin{array}{ccc}
0&0&0\\
0&0&-1\\
0&1&0
\end{array} \right)}
& {L_2 = \left(
\begin{array}{ccc}
0&0&1\\
0&0&0\\
-1&0&0
\end{array} \right)}
& {L_3 = \left(
\begin{array}{ccc}
0&-1&0\\
1&0&0\\
0&0&0
\end{array} \right)}
\end{array}
\end{displaymath}
These generators of rotations in $\R^3$ obey the commutation relations
\begin{displaymath}
[L_i,L_j] = \sum_{k=1}^{3} \epsilon_{ijk} L_k
\end{displaymath}

\subsection{Spin-$\frac{1}{2}$ Representation of $\mathcal{L}(SO(3))$}
Let $\hbar \in \R_{>0}$ be a positive real number.

Transferring directly from the classical case, we define the 2-D unitary representation of $\mathcal{L}(SO(3))$ by the Lie-Algebra homomorphism $\mu$
\begin{displaymath}
\mu(\vec{L}) = -\frac{i}{\hbar} \hat{\vec{S}}
\end{displaymath}
where the $\hat{\vec{S}}$ are our hermitian spin operators $\hat{\vec{S}} := \frac{1}{2}\hbar \hat{\vec{\sigma}}$ acting on $\C^2$, and the Pauli matrices
are the usual
\begin{displaymath}
\begin{array}{ccc}
{\hat{\sigma}_1 = \left(
\begin{array}{cc}
0&1\\
1&0
\end{array} \right)}
& {\hat{\sigma}_2 = \left(
\begin{array}{cc}
0&-i\\
i&0
\end{array} \right)}
& {\hat{\sigma}_3 = \left(
\begin{array}{cc}
1&0\\
0&-1
\end{array} \right)}
\end{array}
\end{displaymath}

These hermitian operators of course obey the induced commutation relations
\begin{displaymath}
[\hat{S}_i,\hat{S}_j] = i\hbar \sum_{k=1}^{3} \epsilon_{ijk} \hat{S}_k
\end{displaymath}

And our chosen basis for $\C^2$ provides an eigenbasis for the hermitian linear operator $\hat{\sigma}_z$, such that the spin-up and spin-down states have
the expected spin angular momentum eigenvalues $\frac{1}{2}\hbar, -\frac{1}{2}\hbar$
\begin{displaymath}
\begin{split}
\hat{\sigma}_z \ket{\uparrow} &= \phantom{-}\ket{\uparrow} \\
\hat{\sigma}_z \ket{\downarrow} &= -\ket{\downarrow}
\end{split}
\end{displaymath}

\subsection{Two electrons}
As an example, to model a pair of electrons' spin interaction, we use the tensor product $\C^2 \otimes \C^2$ as the combined spin state-space for the two
electrons, and a simple time-independent Hamiltonian including a spin-spin magnetic interaction proportional to some positive constant $\alpha$.
\begin{displaymath}
\hat{H} = E_1 \hat{\mathbf{1}} + E_2 \hat{\mathbf{1}} -\alpha \hat{\vec{S}}_1 \cdot \hat{\vec{S}}_2
\end{displaymath}

The following aspects of the model are verifiable by direct algebraic manipulation of constant numbers, and are identical to those of the usual
set-theoretic model. The energy and total spin ($\vec{S} = \vec{S}_1 + \vec{S}_2$) eigenstates are
\begin{displaymath}
\begin{array}{llll}
\text{Eigenvector} & \text{Energy} & \vec{S}^2 & S_z\\
\ket{\uparrow}\ket{\uparrow} & E_1 + E_2 - \frac{\alpha\hbar}{2} & 2\hbar^2 & \hbar\\
\ket{\uparrow}\ket{\downarrow} + \ket{\downarrow}\ket{\uparrow} & E_1 + E_2 - \frac{\alpha\hbar}{2} & 2\hbar^2 & 0\\
\ket{\downarrow}\ket{\downarrow} & E_1 + E_2 - \frac{\alpha\hbar}{2} & 2\hbar^2 & -\hbar\\
\ket{\uparrow}\ket{\downarrow} - \ket{\downarrow}\ket{\uparrow} & E_1 + E_2 + \frac{3\alpha\hbar}{2} & 0 & 0\\
\end{array}
\end{displaymath}

The time-evolution of the state (time modelled as $\R$) is subject to the Schr\"odinger equation
\begin{displaymath}
i\hbar \frac{d}{d t} \ket{\psi}_t = \hat{H}\ket{\psi}_t
\end{displaymath}
Or equivalently, in SDG notation
\begin{displaymath}
\ket{\psi}_{t+dt} = \ket{\psi}_t -i \frac{dt}{\hbar} \hat{H}\ket{\psi}_t \bigspace \bigspace (\forall t \in \R, dt \in \D)
\end{displaymath}

The following theorem on synthetic Lie groups is proved in \cite{LieIntegration}.
\begin{theorem}
For any Lie Group $G$, with Lie Algebra $\mathcal{L}G$ we have
\begin{displaymath}
(\forall f \in \mathcal{L}G^{\R}) \Big(\exists ! F \in G^{\R} \text{ with } F(0) = e \text{ and } (\forall t \in \R, d \in \D)(F(t+d).F(t)^{-1} =
f(t)(d))\Big)
\end{displaymath}
\end{theorem}

This theorem is the analogue of the classical integration theorem for Lie groups, and states the existence of a unique time-evolution for a
finite-dimensional quantum system, given even \emph{any} time-dependent Hamiltonian. To see this, let $f(t)$ be the time-dependent Hamiltonian, expressed
as a varying element of the Lie Algebra $\mathcal{L}U(n)$ for the unitary group $U(n)$. Then the function $F(t)$ provides the evolution operator (an
element of $U(n)$) for time $t$. For a time-independent Hamiltonian ($f(t)$ independent of $t$), the exponential map
\begin{displaymath}
\exp : \mathcal{L}U(n) \maps U(n)
\end{displaymath}
is simply $\exp (f) = F(1)$.

In particular then, the above Hamiltonian suffices to uniquely determine the time evolution of the joint quantum state, as in the classical case. The
evolution is of course expressible in terms of the coefficients of the state vector in the energy eigenbasis, using the complex exponential function.

\subsection{Measurements and an Interpretation of Probability}

For any given quantum-mechanical system (state-space, initial state, and Hamiltonian), then it is a magical and convenient fact that this, plus the
specification of a measurement at a given time, is enough to reduce the space of outcome probabilities to that of the familiar Cauchy reals $\SetsR_C$,
even though the inner-product has as its codomain the object of smooth reals $\R$. It is therefore no good to go looking for differences in probabilistic
predictions between the two theories in any \emph{given} situation, because they simply will not be found.

The underlying reasons for this can be understood entirely within topos theory. In topos theory, a \emph{global} element $x$ of an object $X$ is a
morphism from the terminal object of the topos, $\terminal$, to the object $X$. This parallels the set-theoretic case of an element $y$ of a set $Y$ being
equivalent to a function from the one-element set $\{*\}$ to the set $Y$. Now in set theory, all elements of an object are of this exact form, but not so
in topos theory. There are other types of element, not necessarily global in nature. It is the global elements that in the internal logic of the topos can
be treated as precisely the elements that are capable of being \emph{named}, or pinpointed, or that if one prefers, are the elements that can be imagined
as separate from all other elements.
    So a global element $x$ of an object $X$ is a morphism
\begin{displaymath}
\terminal \xrightarrow{x} X
\end{displaymath}
In topos theory, as in set theory, a formula $\phi$ with domain $X$ is a morphism from $X$ to the object of truth-values $\Omega$.
\begin{displaymath}
X \xrightarrow{\phi} \Omega
\end{displaymath}
One may think of the formula $\phi$ as a yes/no question about each element of $X$, and the $\phi$-image of a particular element $x$ as the answer to the
question for $x$. This view is central in both set theory and topos theory to the building of sets by the Principle of Comprehension\footnote{In set
theory, given any formula $\phi: X \maps \{true, false\}$, the Principle of Comprehension is an axiom stating that we may construct the subset $\{x \in X
\lsuchthat \phi(x)\}$. In topos theory, a generalisation of the axiom to the use of a different object of truth values, $\Omega$, is demanded as a part of
the definition of a topos.}. Now if one asks a question of a global element of an object, one will obtain a global element of $\Omega$ as the answer. In
set theory, and topos theory, this process is represented by the composition of the maps representing the element and the formula.
\begin{displaymath}
\begin{split}
\terminal \xrightarrow{x} X \xrightarrow{\phi} \Omega \\
\terminal \xrightarrow{\phi \of x} \Omega \\
\end{split}
\end{displaymath}
In both topos theory and set theory, $\phi \of x$ is usually written $\phi(x)$. Of course, when $x$ is a global element of $X$, one sees that $\phi(x)$ is
a global element of $\Omega$. The more logically closely related toposes to $\Sets$ have only two global truth values, $true: \terminal \maps \Omega$ and
$false: \terminal \maps \Omega$ (though generally this does \emph{not} force $\Omega \cong \{true, false\}$). All of the well-adapted models of SDG
possess this quality, and therefore give yes/no answers to all individual questions asked of \emph{given} objects and elements. This seems to contradict
the tenet that the law of excluded middle is not universally valid. The subtlety is in the condition of explicitly naming the element we are testing the
proposition for. The explicit specification of the element forces the truth-value of any proposition applied to the element to also be explicitly
specifiable. In this way, a specified Hilbert space, initial state, time of measurement, and measurement will give global elements of $\R$ as the
probabilities for the outcomes of a measurement. But the global elements of $\R$ are precisely the elements of $\SetsR_C$, and such values of probability
can be interpreted in a relative frequency way, for an $\SetsN$-indexed number of identical measurements. One will only struggle to find an interpretation
for $\R$ as a space of probabilities if one actually performs either a logically ill-defined experiment (one that is not explicitly describable), or an
infinite sequence of only implicitly described experiments (e.g. experiments of the form ``for each $\theta \in \D$ rotate an electron by $\theta$ radians
about the x-axis, and measure the spin in the z-direction"), which so far as we know is impossible!

\subsection{Numerics}

If one of the aims of this paper is to persuade theoretical physicists that the use of a more general logic is not something so outlandish as it might at
first seem, then it is worth commenting on the numerics of $\R$. As an example: the relationship between a decimal representation and a real number.

We have already shown that no projection map from $\R$ to $\SetsR_C$ can exist in any universe containing SDG, and therefore there is no general method
(represented by such a function) for the approximation of a real number by an infinite decimal expansion (the Cauchy reals are effectively defined as the
set of decimal expansions). An $n$-decimal-place expansion $x_n$ is a valid approximation to a real number $x$ if $\abs{x-x_n} \leq 0.5 \times 10^{-n}$.
The fact that the sequence of errors $0.5 \times 10^{-n}$ tends uniquely to zero in $\SetsR_C$ is used to allow a decimal expansion to pinpoint a Cauchy
number. However, considered as a sequence of errors in $\R$, this also tends to all elements of the object of infinitesimals $\infinitesimals := \{x\in
\R|(\forall n\in \SetsN)\frac{-1}{n+1}<x<\frac{1}{n+1}\}$, and so does not uniquely pinpoint any member of $\infinitesimals$.
\begin{displaymath}
\begin{split}
x_n &\rightarrow x\\
&\Downarrow \\
x_n &\rightarrow x + \infinitesimals
\end{split}
\end{displaymath}

The topological real line axiom (\ref{axiom:OPENCOVER}) gives us a cover of $\R$ that is a starting point for the process of obtaining a decimal
expansion. \setcounter{axiom}{4}
\begin{axiom}[Open Cover]
\begin{displaymath}
(\forall x \in \R)(x<1 \lor x>0)
\end{displaymath}
\end{axiom}

If $\R$ is then taken to be Archimedean\footnote{In the more advanced models (e.g. $\B$, $\Z$ of \cite{Models}) of SDG $\R$ is not Archimedean, but then
either one can extend the `countable' cover to an `s-countable' cover involving the use of an alternative natural numbers candidate, or just consider the
arithmetically \emph{accessible} part of $\R$ instead of the whole of $\R$.} (i.e. ($\forall x \in \R)(\exists n \in \SetsN) (x<n)$), the previous
cover-axiom can be used to provide the family of countable covers $B_n(p,q) := \Big((n-0.5) \times 10^{-p} - \frac{1}{q}, (n+0.5) \times 10^{-p}\Big)$
(where $p$ and $q$ are natural number valued accuracy and overlap values parametrising the family).

One might imagine that these covers are suitable food for the creation of a method for generating decimal expansions, but this is only true for
individually imagined real numbers. Given any accuracy and overlap parameters, the extension to a general method (for all numbers) involves either the
application of excluded middle or the axiom of choice, one of which must be invoked to choose an open interval for each of the real numbers one in
principle considers\footnote{Many numbers lie in two open intervals. Many lie in just one. To decide between these two cases requires the excluded middle.
If no explicit decision of this form is considered necessary then we require the axiom of choice or Zorn's Lemma to proceed.}. The axiom of choice does
not hold in non-boolean toposes; a failure that helps allow the generalisation from set theory.

Just because a general method for producing decimal expansions does not exist, it does not stop us from thinking about smooth real numbers using this
familiar concept, so long as we are aware that infinitesimal but significant errors can crop up in doing so, and these might lead to logical
contradictions and misunderstanding. The name of a real number is equivalent to its decimal expansion, but not all numbers are nameable. Physicists might
naturally ask about the representation of the other elements of the real line using decimal expansions, but our point is that we do not need to have one,
since the input parameters and results of any physics calculation will be nameable numbers. The trick is to use decimals only to input empirical data and
to read off the results of calculations. Decimals have no place in the intermediate stages.

\section{Discussion}

\subsection{Unnormalisable State Vectors}
It is an interesting question to consider the physical meaning of unnormalisable vectors of the form $\epsilon\ket{\psi}$ where $\epsilon$ is a nilpotent
real. These are perfectly reasonable vectors in the context of synthetic differential geometry. As an example, take the vectors $d\ket{\uparrow}$ for $d
\in \D$, where $\D$ is the object of square-zero reals. One might expect that if such vectors were able to fully describe some of the quantum states of a
physical system, then the vectors $d\ket{\uparrow} \otimes d\ket{\uparrow}$ would completely describe some of the quantum states comprised of two
identical copies of that system. This cannot be the case, since the latter vector is identically zero, and therefore does not represent a physical state
at all. This suggests that either none of the $d\ket{\uparrow}$ represents any physical quantum state, or that they represent physical states with the
interesting property of not being able to be logically duplicated (or, more generally, considered to be part of a larger quantum system). Intuitionistic
logic tells us that it cannot be shown for any $d \in \D$ that either $d = 0$ or that $d \neq 0$, suggesting that it cannot be shown that either
$d\ket{\psi}$ does or does not represents a physical quantum state. A possible sensible definition of what constitutes a representative of a physical
quantum state follows.
\begin{definition}
A vector $\ket{\psi}$ represents a physical quantum state if and only if $\ip{\psi}{\psi}$ is an invertible real number.
\end{definition}
This coincides with $\ket{\psi} \neq 0$, due to the field axiom (\ref{axiom:FIELD}). Even so, it is not true that all vectors then either do or do not
represent physical states. This is a direct consequence of the need for the lack of validity of the law of excluded middle when modelling the nilpotent
infinitesimals of synthetic differential geometry. One may take this as a purely mathematical statement, and consider only those vectors that do represent
physical quantum states when considering physics. This is equivalent to accepting that an element of $\C P^n$ is the true representative of a physical
state, and noticing that no ray of Hilbert space is uniquely defined by any vector $\ket{\psi}$ for which we cannot establish $\ket{\psi} \neq 0$ (for any
ray $\rho$, we cannot say that $\ket{\psi} \notin \rho$).

In the unlikely event that the definition above is not accepted, then to what extent the `almost zero' vectors represent physical states is an unexplored
area. Its study may be useful when considering the states of open quantum systems (systems with states that perhaps, even logically, should not be
duplicated by tensor production). It may force upon us the strange and exciting position of being able to talk about the entanglement properties of a
system while having to accept that the system itself neither exists nor does not exist, but instead lies somewhere inbetween. There is however another
strong physical case for the acceptance of the above definition. For any two `nilpotent length' vectors $\epsilon_1\ket{\psi}$ and $\epsilon_2\ket{\phi}$,
we are prevented from asserting the logical statement $\epsilon_1\ket{\psi} \neq \epsilon_2\ket{\phi}$. This fact logically implies that no physical
difference can be found between states represented by the two vectors; for if one could, we would have a method to decidedly tell apart the two vectors.
This is strongly related to the non-uniqueness of ray generated by the vectors. This is not to say that future developments in logic and set theory will
not be able to consider this possibility.

\subsection{Infinite-Dimensional Representations}
It is natural to ask about the extension to an infinite number of dimensions of quantum state-space. There are some problems associated with this
procedure, in defining and using the inner-product, and related to the incompleteness of the real line $\R$. Restricting a vector space to only the span
of some (Hamel) basis would certainly preserve the synthetic smoothness of the inner-product. If however, we allow more general combinations of basis
vectors (e.g. limits of finite combinations) then in passing to an infinite number of dimensions it seems the values of the inner-product could no longer
belong just to $\C$. The question of whether this smoothness is desirable is a question for the probability aspects of the theory, as probabilities are
conventionally calculated using the inner-product.

Progress has been made on the building of suitably defined Banach spaces over the field $\SetsR_D$ of Dedekind reals in toposes~\cite{BanachSheaves}. This
involves a modification of the definition of a Cauchy sequence to render complete the also previously sequentially incomplete $\SetsR_D$. Not only does
such a modification fail on the object of smooth reals $\R$ (which is a sub-object of $\SetsR_D$ in at least $\F$ and $\G$), but even if one were able to
show in some similar scheme that $\R$ could be regarded as complete, such a scheme would still reflect a severe restriction on the allowed combinations of
vectors whose inner-products are smooth complex numbers, as compared to the set-theoretic case. In set theory, the vector space $\{\vec{x} \in
\SetsC^{\SetsN} \lsuchthat \phi(\vec{x})\}$ (where $\phi(\vec{x})$ is the formula ``$\sum_{i=1}^{\infty}\bar{x_i}x_i \in \SetsR$'') is a Hilbert space
under the inner-product $\vec{x}\cdot\vec{y} = \sum_{i=1}^{\infty}\bar{x_i}y_i$. To see that the inner-product is well-defined and has codomain $\SetsC$,
one must show that $\sum_{i=1}^{\infty}\bar{x_i}y_i$ converges to a unique complex number for any pair $(\vec{x}$, $\vec{y})$ whose squares converge to
real numbers. This makes use of the Cauchy-Schwartz inequality and the sequential completeness of $\SetsR$. However, the formula $\phi$ may not be a
strong enough condition (in topos theory) to retain $\C$ as the codomain of the inner-product when the transition from a finite to an infinite number of
dimensions is made. To decide whether $\phi$ is a strong enough condition happens to be a non-trivial mathematical question. If $\phi$ is not strong
enough, this would suggest either a difference in the behaviour of the probabilities generated by finite and infinite-dimensional state-spaces, or a new
logical restriction in the allowed physical states of an infinite-dimensional quantum system.

\subsection{With Hindsight} Leibniz was primarily interested in law and logic, being a philosopher and a lawyer as well as an accomplished mathematician.
It is important to realise that his pseudo-logical remarks on the nature of infinitesimals did not come entirely from attempts to defend his use of them,
but also from a strong instinct for logic. The subobject of the infinitesimal numbers most useful to differential calculus is the object of nilpotent
infinitesimals $\D_{\infty}$. Assuming with hindsight that Leibniz was only in effect ever considering the nilpotent infinitesimals, it is interesting to
compare some of his comments with what SDG has to say about them.

``An infinitesimal is less that any given quantity."

\begin{displaymath}
SDG \models (\forall i \in \D_{ \infty })(\forall x \in \SetsR_C^{>0})(i<x)
\end{displaymath}

``[Infinitesimals are] qualitative zeroes.''

\begin{displaymath}
SDG \models (\forall d \in \D_{\infty})( \lnot d \neq 0)
\end{displaymath}

Leibniz stated that he ``did not believe at all that there are magnitudes truly infinite or truly infinitesimal''.

\begin{displaymath}
\begin{split}
SDG \models \lnot (\exists d \in \D_\infty)(d \neq 0) \\
SDG \models (\forall d \in \D_{\infty})(\not \exists d^{-1})
\end{split}
\end{displaymath}

Leibniz last of all proposed the existence of infinitesimals only as ``auxiliary variables''. This is the analogue in SDG of the need to consider such
things as the range $\forall d \in \D$ rather than any individual member of $\D$ when writing down differential formulae (c.f. Kock-Lawvere axiom).

With the development of topos theory and (within) synthetic differential geometry, we have in hindsight seen that the path of mathematical thinking taken
by humankind has been a choice left to chance by the assumptions of the early pioneers. Without realising that a plethora of alternatives to a standard
way of thinking exists, mathematicians have developed their theories in a minute part of the logical multiverse that has now been constructed. Looking
back to the work of the two founders of calculus, Newton and Leibniz, we can see that they were in fact developing their maths in different universes.
Newton unwittingly used set theory, and Leibniz topos theory. That they were not aware of this fact is down to the subtlety in that difference, and the
lack of good foundations for the effective set theory of the time. Leibniz's universe was rejected after a forceful development of the new accessible idea
of limits was made, in the relief that a tangle of logic had been resolved. The results of this path more closely agreed with Newton's explanations. It
cannot be stressed enough that this choice of mathematical universe must have had a huge effect on the method of approach to many problems over the
centuries. The understanding of many other mathematical universes requires an appreciation of topos theory, whereas the understanding of the currently
default mathematical universe only requires the appreciation of the relatively simple set theory. That books on topos theory develop it using set theory
is not something that should be seen as a reason for the rejection of intuitionistic universes as somehow less real. That pen-and-paper notation (or
language) is required to communicate ideas between people is at the heart of why set theory is easier maths than is topos theory, and why topos theory
appears to rest on set theory. Set theory is for the most part the study of the manipulation of collections of ink symbols marked on paper, and it is the
human assumption that these markings are sufficient for directly representing all mathematical objects that lead to its use as the first category
supposedly directly modelling all mathematical visualisations. The logic of manipulating these symbols is of course then projected onto the workings of
the mathematical universe they are supposed to represent. That topos theory can access through pen-and-paper notation a qualitatively larger universe of
ideas is a wonder, and necessarily a more complicated affair than the direct representation of sets and their elements by single bits of ink on paper.
Further, it is a strange thing that Leibniz had such a very strong instinct for the workings of such universes.

If human mathematical thought had instead taken another route, and developed along the lines of intuitionistic logic instead of restricting to classical
logic, it would be interesting to see what the current approaches to the problems of physics might have been. The current approach to foundational physics
tends more and more towards discrete ideas in mathematics (causal sets, holographic principles), which coincides with the foundational approach to
set-theoretic maths. It can be imagined that if there are real differences in the logical modelling of a continuum, then there could be real differences
in the method by which each could arise from more discrete predecessor (causal sets $\maps$ Cauchy spacetime, ? $\maps$ smooth spacetime). Parallels might
have been drawn far earlier between, say, the method by which the physical universe always hides virtual particles from the observer, and the method by
which a mathematical universe can hide infinitesimals from the man who wishes to name one.

\section{Conclusions}

We have argued that the new logical perspective afforded by topos theory is suitable to the task of describing the physical world around us, at least as
well as is set theory. That doing so requires more thought and a better appreciation of the assumptions going into any physical model was demonstrated by
a walk-through of some of the maths chosen to model a simple quantum system. This increased labour of thought we argue might be enough to provide
significant insights into current and future approaches to foundational physics.

We noted on our foray into quantum mechanics such observations as the fact that complex Hilbert spaces and finite-dimensional complex inner-product spaces
need not be the same thing when one frees oneself from classical logic, and that the extension of simple quantum mechanics from finite to
infinite-dimensional systems seems to require either a change in the behaviour of probability, or a change in the collection of allowed physical states.

We claim that the early human decision to restrict to set-theoretic thinking has resulted in an unduly narrow perspective of physics, particularly in its
foundations, and that this decision was not a well-informed one, despite having been tremendously useful to date.

\section{Acknowledgements}

Many thanks to Chris Isham for patient perseverance during the long period of struggle freeing myself from the once ingrained belief that set theory
describes all, and of course many useful physics discussions during the writing of this paper. Also to Peter Johnstone for his patience and extremely
helpful answers to a physicist's questions on topos theory.

\newpage

\begin{appendix}
\section{Intuitionistic Logic and Toposes}
\subsection{Proof By Contradiction}
Formal intuitionistic logic (IL) was pioneered by Arend Heyting early last century, and supplies a logical framework for \emph{constructive} maths.
Constructivists reject as a fundamental axiom the law of excluded middle, which allows proof by contradiction. Using the symbols $\ltrue$, $\lfalse$,
$\limplies$, $\lnot$ and $\lor$ to mean \emph{true, false, implies, not} and \emph{or} respectively, we say that a statement $\phi$ is not true, or write
$\lnot \phi$, if the assumption of $\phi$ leads to a contradiction, i.e. if $\phi \limplies \lfalse$. Thus a proof of the negation of a statement can be
accomplished in a literal sense by a derivation of contradiction. However, this is not what is meant by the term proof by contradiction. Proof by
contradiction, by convention, is the more complicated series of steps starting with the assumption that we wish to prove a statement $\psi$ by deriving a
contradiction from the supposition of its negation $\lnot \psi$ and marrying that conditional contradiction with the axiom $\psi \lor \lnot \psi$ to
finally obtain $\psi$ itself. We thus prove $\psi$ by contradiction, a process that can be succinctly represented by the logical consequence $\lnot \lnot
\psi \liff \psi$ of the more basic $\psi \lor \lnot \psi$.
\subsection{IL}
IL does not assume that the law of excluded middle does not apply to anything, it merely demotes the universal nature of the law to a statement about
statements that itself can assume different truth values in varying circumstances, thus \emph{generalising} classical logic (CL). The following eleven
axioms remain universally true in propositional IL, however.
\begin{displaymath}
\begin{split}
\alpha &\limplies (\alpha \land \alpha) \\
(\alpha \land \beta) &\limplies (\beta \land \alpha) \\
(\alpha \limplies \beta ) &\limplies ((\alpha \land \gamma ) \limplies (\beta \land \gamma )) \\
((\alpha \limplies \beta) \land (\beta \limplies \gamma)) &\limplies (\alpha \limplies \gamma) \\
\beta &\limplies (\alpha \limplies \beta) \\
(\alpha \land (\alpha \limplies \beta)) &\limplies \beta \\
\alpha &\limplies (\alpha \lor \beta) \\
(\alpha \lor \beta) &\limplies (\beta \lor \alpha) \\
((\alpha \limplies \gamma) \land (\beta \limplies \gamma)) &\limplies ((\alpha \lor \beta) \limplies \gamma) \\
\lnot \alpha &\limplies (\alpha \limplies \beta) \\
((\alpha \limplies \beta) \land (\alpha \limplies \lnot \beta)) &\limplies \lnot \alpha \\
\end{split}
\end{displaymath}

\subsection{Toposes}
A topos is a category $\topos$ with the following properties.
\begin{displaymath}
\begin{split}
&\topos \text{ has a terminal object } \terminal \text{ and pullbacks} \\
&\topos \text{ has an initial object } \initial, \text{ and pushouts} \\
&\topos \text{ has exponentiation } \\
&\topos \text{ has a subobject classifier } \ltrue : \terminal \maps \truthvalues
\end{split}
\end{displaymath}
The prime and motivating example is the category of $\Sets$. The conditions above allow the category $\topos$ to be used as a model of a particular IL
universe in much the same way as $\Sets$ is used as a model of a CL universe. A few examples of the categorical structures representing logical ideas are
the following.

\textbf{E.g. 1)} The object of truth values is labelled $\truthvalues$. In $\Sets$ this is simply $\{ true, false \}$.

\textbf{E.g. 2)} A subobject\footnote{See any book on topos theory for a more precise definition involving equivalence classes of arrows.} $a$ of $X$ is a
monic arrow $a: A \monic X$. In $\Sets$ this is simply a subset $A$ of a set $X$.

\textbf{E.g. 3)} The arrow $\lnot : \truthvalues \maps \truthvalues$ representing logical negation is given as the character of the arrow $\lfalse:
\terminal \maps \truthvalues$, which in turn is defined as the character of the (unique) arrow $\initial \maps \terminal$.

For an excellent introduction to topos theory, see \cite{Goldblatt}.

\section{Sheaf Categories as Toposes}
Many toposes are categories of sheaves. A sheaf is a contravariant functor from a \emph{site} $\SetsC$ to the category of sets $\Sets$, that respects the
topological structure of the site. A site $\SetsC$ is a category of `areas' with topological structure defined by the requirement that each area $\lA$ in
$\SetsC$ possesses a family of covers. These covers must obey the following conditions.
\begin{displaymath}
\begin{array}{l}
\text{i) The single-element set } \{\lA \xrightarrow{id_{\lA}} \lA\} \text{ is a cover.}\\
\text{ii) If } \{\lX_i \xrightarrow{x_i} \lA \}_i \text{ covers } \lA \text{, and if for each } \lX_i \text{ we have that } \{\lY_{ji}
\xrightarrow{y_{ji}}
\lX_i\}_j \text{ covers } \lX_i \text{, then }\\
\phantom{ \text{ii) If }} \{ \lY_{ji} \xrightarrow{y_{ji}} \lX_i \xrightarrow{x_i} \lA \}_{ji} \text{ also covers } \lA .\\
\text{iii) If } \{\lX_i \xrightarrow{x_i} \lA \}_i \text{ covers } \lA \text{, then for any arrow } \lB \xrightarrow{b} \lA\\
\phantom{\text{iii) If }} \{ \lB \times_{\lA} \lX_i \xrightarrow{x_{i}|_b} \lB \}_i \text { is a cover of } \lB.\\
\phantom{\text{iii) Marvellous: }} \text{(}\times_{\lA} \text{ and } x_{i}|_b \text{ are categorical pullback notation.)}
\end{array}
\end{displaymath}

We say that a contravariant functor $F: \SetsC \maps \Sets$ respects the site's topological structure if it satisfies the following condition.
\newline
\newline
For any cover $\{\lX_i \xrightarrow{x_i} \lA \}_i$, and any collection of elements $f_i \in F(\lX_i)$ that are pairwise compatible\footnote{i.e. They must
agree on the intersection of the areas. The intersection of two areas of a site is required to exist as their categorical pullback.} in the functorial
sense, there is a unique $f \in F(\lX)$ such that $(\forall i)(f_i = F(x_i)(f))$.
\newline

A sheaf morphism is a natural transformation between two sheaves.

An example of a site is the collection of open subsets of $[0,1]$, and all inclusion maps. Its areas are the open subsets of $[0,1]$, and for each open
subset, its covers are the open covers of that subset. A sheaf over $[0,1]$ is usually referred to as a sheaf of continuous sections over $[0,1]$.

The site, and its topological structure, wholly determines how logical information in the topos is organised and how it flows. Of course, the idea that
logical information flows from area to area might sound a little alien to set-theorists. This is because any set may be viewed as a sheaf over the single
element site $\terminal$. In set theory, the logical information may be in only one place, and cannot flow anywhere; it is static. The ultimate idea is
that the topos itself does not `know' of the site over which its objects are constructed, or of any flow of logic. Nevertheless, as a tool to
understanding, the site tells us exactly how the logic of the topos does work. We said that a sheaf was a functor respecting the topological structure of
the site over which it is defined. In this way, a sheaf represents a generalised set respecting the logic of the mathematical universe in which it
resides.

A sheaf $X$ over the site $\SetsC$ can be viewed as a rule that associates to each `logical area' $\lA$ of the site a view of $X$ (a set): $X(\lA)$,
imagined as the way the logical area `sees' the object $X$.

A sheaf morphism is then, for each area, essentially just a map from its view of one sheaf to its view of the other, in such a way that the morphism
components commute with view restrictions. In this way it is to be imagined that a generalised function is a map from one generalised set to another that
preserves or respects the flow of internal logic of the universe. Sheaves and sheaf-morphisms simply reduce to sets and maps between sets when the site
over which sheaves are defined is the single element site $\terminal$, with its unique topological structure.

\subsection{Number Systems in SDG}

\subsubsection{The Naturals, Rationals and Cauchy Reals}
The naturals, $\SetsN$, in any topos that possesses them, can be thought of in exactly the same way as the naturals we are all familiar with. Showing this
is another matter, and is left to texts such as \cite{Johnstone}. From the natural numbers $\SetsN$, the integers $\SetsZ$ and rationals $\SetsQ$ can be
easily constructed. These objects also behave as expected. Classical reasoning is entirely permissible when considering the arithmetic of these number
systems, as they are all decidable objects. So too is the object of Cauchy reals $\SetsR_C$, constructed as equivalence classes of sequences of rationals.

As an example, the topos of sheaves over $[0,1]$ has a natural numbers object. It is given by
\begin{displaymath}
\begin{array}{l}
\SetsN(U) := \{f:U \maps \SetsN\ \lsuchthat f \text{ is continuous}\}\\
\SetsN(V \xrightarrow{i} U)(f) := f|_V
\end{array}
\end{displaymath}

Note that in the topos of sheaves over $[a,b]$, where $a=b$, the natural numbers object is the same as that of set theory.

\subsubsection{The Smooth Reals}

The smooth real object was discovered inhabiting many toposes. Some of its properties were abstracted to provide the foundation for synthetic differential
geometry, however the smooth real object of one topos generally behaves slightly differently to those of others. The axiomatic scheme we assume for the
smooth reals is as follows. $\R$ is to model the axioms (A1) - (A15) proposed by I.~Moerdijk and G.E.~Reyes, listed on pages 295-298 of their
work~\cite{Models}. We use $U(\R)$ to denote the subobject of invertible elements. The most relevant of those axioms for this paper are the following.
\setcounter{axiom}{0}
\begin{axiom}[A1] \label{axiom:A1} $\R$ is a commutative ring with unit.
\end{axiom}

\begin{axiom}[A2] \label{axiom:A2} $\R$ is local, such that
\begin{displaymath}
\begin{split}
&0 \neq 1\\
&(\forall x)\big(x \in U(\R) \lor (1-x) \in U(\R)\big)
\end{split}
\end{displaymath}
\end{axiom}

\begin{axiom}[A3] \label{axiom:A3} $(\R,<)$ is a Euclidean ordered local ring, such that
\begin{displaymath}
\begin{split}
&0<1\\
&(0<x) \land (0<y) \limplies (0 < x+y) \land (0 < xy)\\
&x \in U(\R) \liff 0<x \lor x<0\\
& (0<x) \limplies \exists y (x=y^2)\\
\end{split}
\end{displaymath}
\end{axiom}

We take the following additional field axiom, relating distinguishability and invertibility, to hold true for our ring $\R$.
\begin{axiom}[Field] \label{axiom:FIELD}$\R$ is a field in the following sense
\begin{displaymath}
(\forall x_1,\ldots,x_n \in \R) \Big( \lnot(x_1=0 \land \ldots \land x_n=0) \limplies (x_1 \in U(\R) \lor \ldots \lor x_n \in U(\R))\Big)
\end{displaymath}
\end{axiom}

Also assumed is the following topological axiom
\begin{axiom}[Open Cover] \label{axiom:OPENCOVER}
\begin{displaymath}
(\forall x \in \R) \big( (x<1) \lor (x>0) \big)
\end{displaymath}
\end{axiom}
This says that the object $\{(\leftarrow,1),(0,\rightarrow)\}$ is an open cover.
\newline
\newline
Fundamental to SDG are the differentiation and integration axioms
\begin{axiom}[Kock-Lawvere]
For each $f:\D \maps \R$, there exists a unique $b \in \R$, such that for every $d \in \D$
\begin{displaymath}
f(d)=f(0)+d.b
\end{displaymath}
\end{axiom}

\begin{axiom}[Integration]
For each $f:\R \maps \R$, there exists a unique $F: \R \maps \R$, such that
\begin{displaymath}
\begin{split}
F'(x) &= f(x)\\
F(0) &= 0\\
\end{split}
\end{displaymath}
\end{axiom}

\section{Toposes Modelling SDG}
The well-adapted models of SDG discovered are all Grothendieck toposes (categories of sheaves over sites). Extensive studies of the SDG aspects of these
toposes (and others) have been made by I.~Moerdijk and G.E.~Reyes in their book `Models for Smooth Infinitesimal Analysis'~\cite{Models}. Those readers
who wish to examine the toposes are referred to this source. The studies have revealed some of the variable behaviours of the smooth reals as one moves
from one topos to another. It is inevitable that there will be preferred choices of topos for the construction of quantum mechanics within.

It is found that the most basic topos modelling parts of SDG ($\locisheaves$) does not possess the nicest of properties regarding $\R$. For example, in
this topos, the interval $[0,1]$ is not compact, $\R$ is not a local ring, and many of the axioms of the previous section are not modelled. It is
difficult to imagine this smooth real object modelling the space around us. The toposes $\F$ and $\G$ detailed in \cite{Models} validate all of the axioms
required in the body of this paper.

It is useful to have something explicit to consider when trying to imagine the real numbers used in this paper. For this purpose, a very brief description
of the topos $\locisheaves$ is given.

\subsection{The Topos $\locisheaves$}

The category $\SetsL$ consists of

i) The finitely generated $\CInfty$ rings $(A,B,C, \ldots)$.

ii) The $\CInfty$ ring-homomorphisms between them $(\theta:A \maps B, \ldots)$.

The opposite category $\SetsL^{op}$ (with objects labelled $\lA, \lB, \lC, \ldots$) is imbued with the indiscrete topology to turn it into a site. The
Grothendieck topos described by the sheaves over this site is precisely the category of functors from $\SetsL^{op}$ to $\Sets$. The functors represent
generalised sets (objects of the topos) and the natural transformations between them represent the generalised functions (morphisms between objects of the
topos).

All the usual sets of set theory (and their functions, logic and behaviour) can then be embedded in $\locisheaves$ by the functor $\Delta$ as follows

\begin{displaymath}
\begin{array}{l}
\Delta: \Sets \maps \locisheaves \\
\Delta(X)(\lA) := X \\
\Delta(X)(\theta: \lA \maps \lB) := id_X:X \maps X \\
\end{array}
\end{displaymath}

Many of the objects $\lA, \lB, \ldots$ of the site $\SetsL^{op}$ are the $\CInfty$ rings of smooth functions on set-theoretic manifolds $M_A, M_B, \ldots$
(though not all). For such $\lA$, the functor $\R$ representing the smooth reals is given by
\begin{displaymath}
\begin{array}{l}
\R(\lA) := \{f:M_A \maps \SetsR \lsuchthat f \text{ is smooth}\} \\
\R(g:\lA \maps \lB) := \Big(G: \{y:M_B \maps \SetsR \lsuchthat y \text{ is smooth}\} \maps \{x:M_A \maps \SetsR \lsuchthat x \text{ is smooth}\} \Big) \\
\text{where } G(y)(a) := y(g(a))
\end{array}
\end{displaymath}

A smooth real number $x \in \R$ as seen by area $\lA = \ell \CInfty(\SetsR)$ is therefore a smooth function $x_{\lA}:\SetsR \maps \SetsR$. A Cauchy real
$c \in \SetsR_C$ as viewed by the same area is a constant function $c_{\lA} : \SetsR \maps \SetsR$. In this way, it is easy to see there are `more'
elements of $\R$ than there are of $\SetsR_C$.

\section{Some Notation from Axiomatic SDG}
The natural numbers
\begin{displaymath}
\SetsN
\end{displaymath}
The smooth reals
\begin{displaymath}
\R
\end{displaymath}
The infinitesimals
\begin{displaymath}
\infinitesimals := \{x\in \R|(\forall n\in \SetsN)\Big(\frac{-1}{n+1}<x<\frac{1}{n+1}\Big)\} \mspace{20mu} \subset \mspace{20mu} \R
\end{displaymath}
The first-order infinitesimals
\begin{displaymath}
\D := \{d\in \R|d^2=0\}
\end{displaymath}
The nilpotent infinitesimals
\begin{displaymath}
\D_{\infty} := \{d\in \R| (\exists k \in \SetsN)d^{k+1}=0\}
\end{displaymath}

For an excellent introduction to axiomatic SDG, see \cite{Lavendhomme}.

\end{appendix}
\end{onehalfspacing}

\end{document}